\documentclass[review]{elsarticle}

\usepackage{lineno,hyperref}
\modulolinenumbers[5]

\journal{Journal of \LaTeX\ Templates}
\journal{Physica B, Conference proceedings SCES 2017}
\newcommand{\sro}{Sr$_{4}$Ru$_{3}$O$_{10}$}

\begin{document}
	
\begin{frontmatter}

\title{In-depth study of the $H -T$ phase diagram of \sro\ by magnetization experiments}
%\tnotetext[mytitlenote]{Fully documented templates are available in the elsarticle package on %\href{http://www.ctan.org/tex-archive/macros/latex/contrib/elsarticle}{CTAN}.}

%\author{F. Weickert}

\author[Aaddress,Baddress]{F. Weickert\corref{mycorrespondingauthor}}
\cortext[mycorrespondingauthor]{Corresponding author: weickert@lanl.gov}

\author[Baddress]{L. Civale,}
\author[Baddress]{B. Maiorov}
\author[Baddress]{M. Jaime}
\author[Caddress]{M. B. Salamon}
\author[Daddress]{E. Carleschi}

\author[Daddress]{A. M. Strydom}
\author[Eaddress]{R. Fittipaldi}
\author[Eaddress]{V. Granata}
\author[Eaddress]{A. Vecchione}

\address[Aaddress]{Florida State University, NHMFL, Tallahassee, FL 32310, USA}
\address[Baddress]{Los Alamos National Laboratory, MPA-CMMS, Los Alamos, NM 87545, USA}
%\address[Caddress]{Los Alamos National Laboratory, MPA-CMMS, Los Alamos, NM 87545, USA}
\address[Caddress]{University of Texas at Dallas, Richardson, TX 75080, USA}
\address[Daddress]{University of Johannesburg, Auckland Park 2006, South Africa}
\address[Eaddress]{CNR-SPIN Institute and University of Salerno, I-84084 Fisciano, Italy}

\begin{abstract}
We present magnetization measurements on \sro\ as a function of temperature and magnetic field applied perpendicular to the magnetic easy $c$-axis inside the ferromagnetic phase. Peculiar metamagnetism evolves in \sro\ below the ferromagnetic transition $T_{C}$ as a double step in the magnetization at two critical fields $H_{c1}$ and $H_{c2}$. We map the $H-T$ phase diagram with special focus on the temperature range 50\,K $\le T \le T_{C}$. We find that the critical field $H_{c1}(T)$ connects the field and temperature axes of the phase diagram, whereas the $H_{c2}$ boundary starts at 2.8\,T for the lowest temperatures and ends in a critical endpoint at (1\,T; 80\,K). We conclude from the temperature dependence of the ratio $\frac{Hc1}{Hc2}(T)$ that the double metamagnetic transition is an intrinisc effect of the material and it is not caused by sample stacking faults such as twinning or partial in-plane rotation between layers.
\end{abstract}

\begin{keyword}
%$\texttt{elsarticle.cls}\sep \LaTeX\sep Elsevier \sep template
%$\MSC[2010] 00-01\sep  99-00
strontiumruthenate, ferromagnetism, metamagnetic transition, magnetic anisotropy, phase diagram
\end{keyword}

\end{frontmatter}

%\linenumbers

%\section{The Elsevier article class}

\paragraph{Introduction}
\sro\ is a strongly correlated electron system that orders ferromagnetically below $T_{C}=105$\,K with a magnetic easy axis along the crystallographic $c$ direction.\cite{crawford_02} It shows a metamagnetic (MM) transition for magnetic fields applied perpendicular to $c$. The metamagnetism manifests as double step in the magnetization.\cite{carleschi_14} It is accompanied by a reduction of the magnetic moment, which points to the existence of antisymmetric exchange coupling in the system.\cite{weickert_17} Pronounced anomalies close to the critical fields $H_{c1}$ and $H_{c2}$ of the MM transition are also observed in transport\cite{fobes_07,xu_07,fobes_10}, and thermodynamic properties.\cite{cao_07,gupta_06,schottenhamel_16} A complete understanding of the MM transition in \sro\ is still lacking despite a great deal of experimental effort in recent years.

One open question is whether the double step in the magnetization is an intrinsic property of the system or caused by misaligned crystallographic domains in the measured single crystal, because it was only discovered recently.\cite{carleschi_14,weickert_17} We check this point with a precise estimation of $H_{c1}(T)$ and $H_{c2}(T)$  for geometrical relations which can distinguish between both scenarios. 

\paragraph{Experiment}
We utilize magnetization measurements to obtain the detailed phase diagram for temperatures up to 120\,K and magnetic fields up to 3\,T. The magnetization is measured with a Quantum Design MPMS. Details of the magnetization experiment as well as sample growth and characterization techniques are described elsewhere.\cite{weickert_17} Special care is taken to start each measurement after zero field cooling the sample. 

\begin{figure}
	\includegraphics[width=4in]{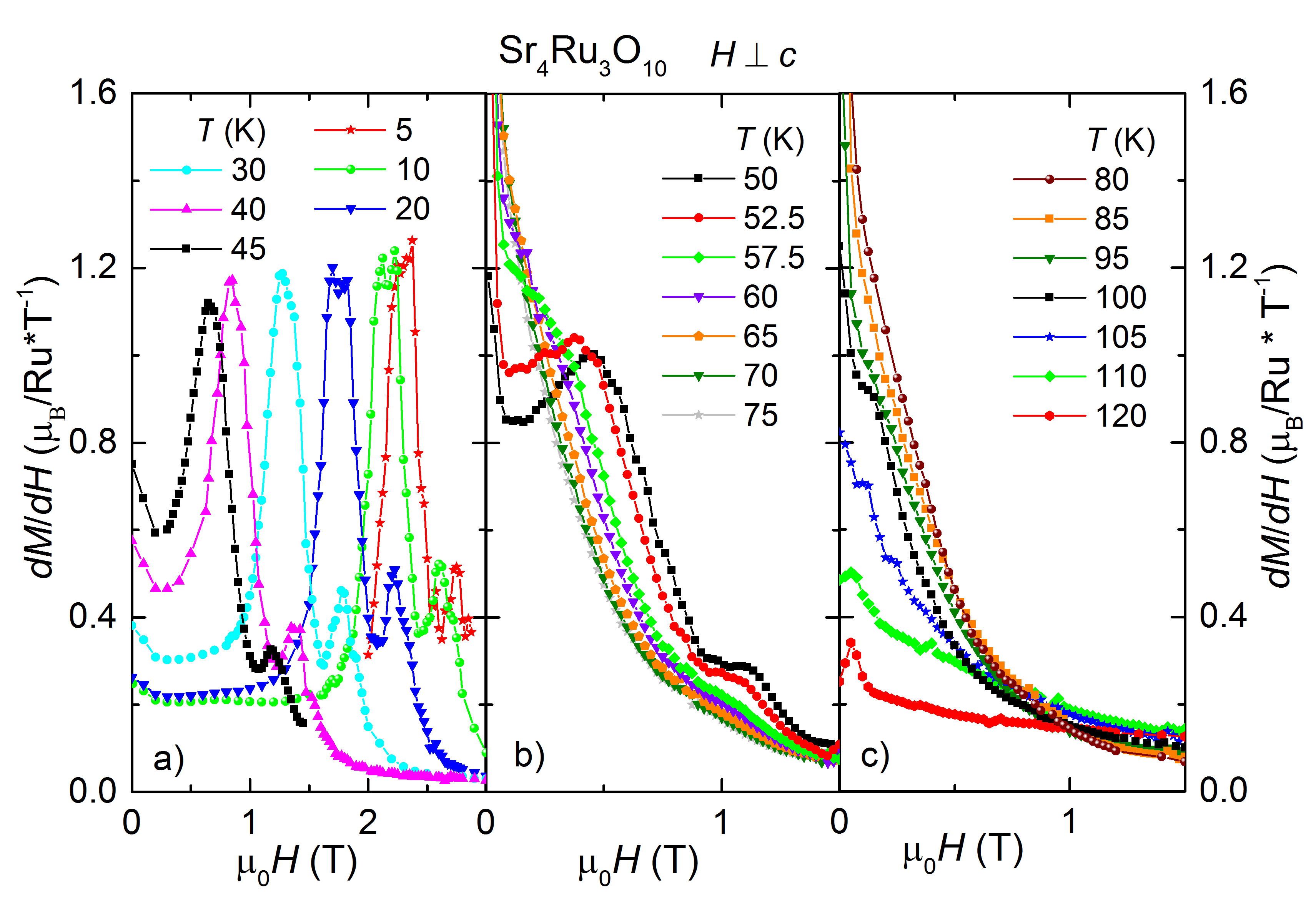}
	\caption{\label{derivative}Magnetic susceptibility $\frac{dM}{dH}(H)$ measured at low temperatures up to 45\,K (a), at intermediate temperatures 50\,K $\le$ 75\,K (b), and at high temperatures 80\,K $\le$ 120\,K (c) of \sro . The experimental points are obtained during increasing magnetic field. We observe two distinct maxima at critical fields $H_{c1}$ and $H_{c2}$ for low temperatures that move to lower field values and vanish above 80\,K. The 85\,K and 95\,K curves shows no features of anomalies and serve therefore as background references in further data analysis.}
\end{figure}

Figure\,\ref{derivative} shows the susceptibility $\frac{dM}{dH}(H)$ for increasing field in 3 different temperature ranges (a) $T <$ 50\,K, (b) 50\,K $\le T \le $ 75\,K, and (c) 80\,K $\le T \le$ 120\,K. A clear double maximum is observed in panel a), accompanied by large hysteresis when compared to measurements on decreasing fields (data not shown). The critical fields $H_{c1}$ and $H_{c2}$ defined at the 2 maxima reach their largest values of 2.3\,T and 2.8\,T below 5\,K. They shift to lower fields in temperatures up to 45\,K, but the sizes of the maximum stay almost constant. The anomalies broaden significantly at intermediate temperatures (panel b) and they vanish completely above 80\,K as shown in panel c.

\begin{figure}
	\includegraphics[width=4in]{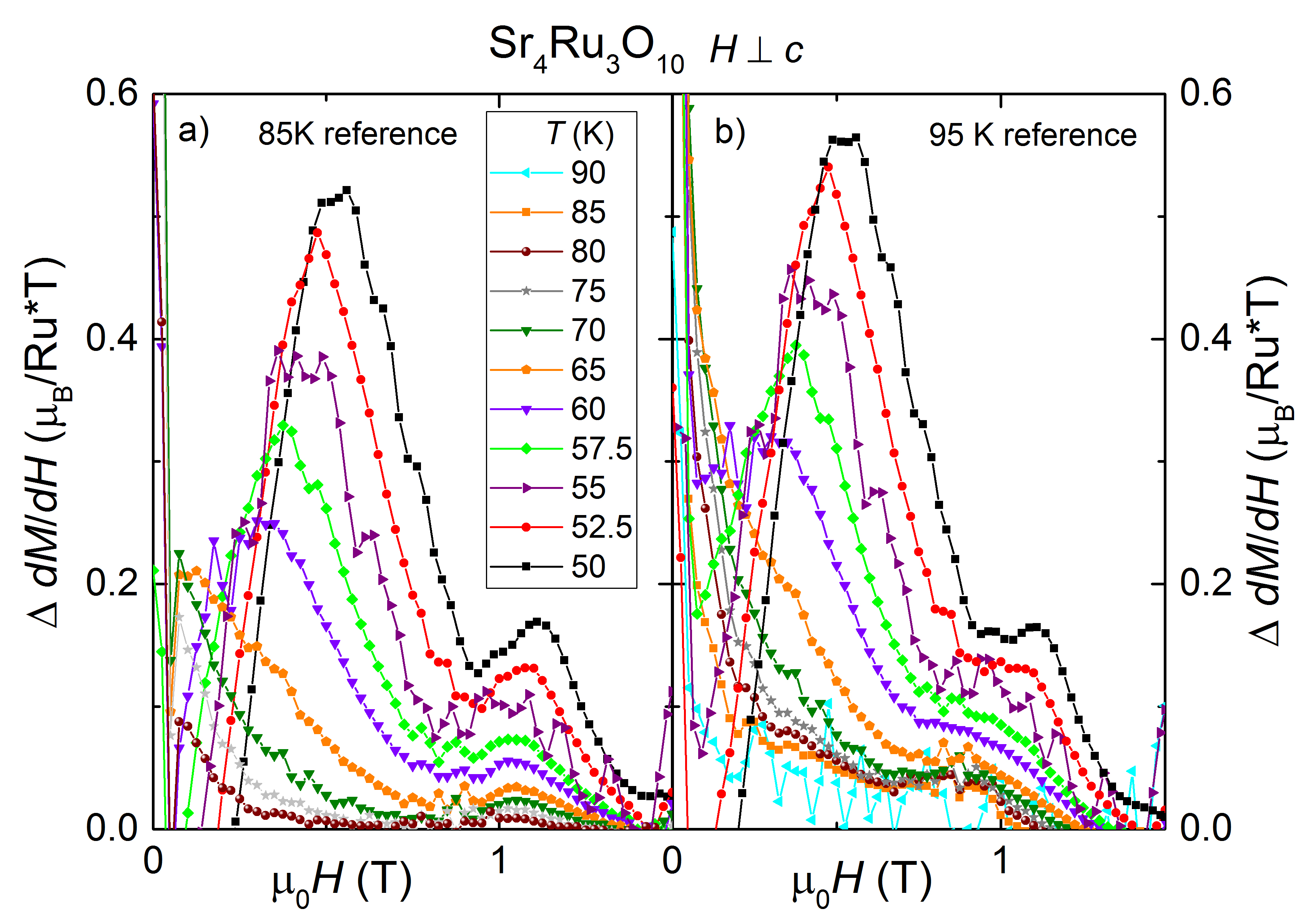}
	\caption{\label{difference_M} Magnetic susceptibility $\Delta \frac{dM}{dH}(H)$ after subtraction of a featureless background (measurement at 85\,K in a) and at 95\,K in b)) in the temperatur range between 50\,K and 90\,K. All curves show a double maximum at $H_{c1}$ and $H_{c2}$, which decreases with increasing temperature and vanishes above 80\,K.}
\end{figure}

We use background subtraction to help identify the MM anomalies at $H_{c1}$ and $H_{c2}$. The measurements taken at 85\,K and 95\,K serve as background (BG). The refined susceptibility $\Delta \frac{dM}{dH} = \frac{dM}{dH}(H,T) - \frac{dM}{dM}(H,T_{BG})$ is displayed in Fig.\,\ref{difference_M} a) for the 85\,T reference and b) for the 95\,T reference.
The position of the maxima in $\Delta \frac{dM}{dH}$ can be tracked in both sets of data all the way up to 80\,K. We observe that $H_{c1}$ moves to zero field whereas $H_{c2}$ shifts only down to 1\,T at high temperatures. There is a small difference in the position of $H_{c1}$ and $H_{c2}$ depending on the chosen background of 85\,K or 95\,K, however, the overall shape of the $H_{c1}$ and $H_{c2}$ phase boundaries are the same.

\begin{figure}
	\includegraphics[width=4in]{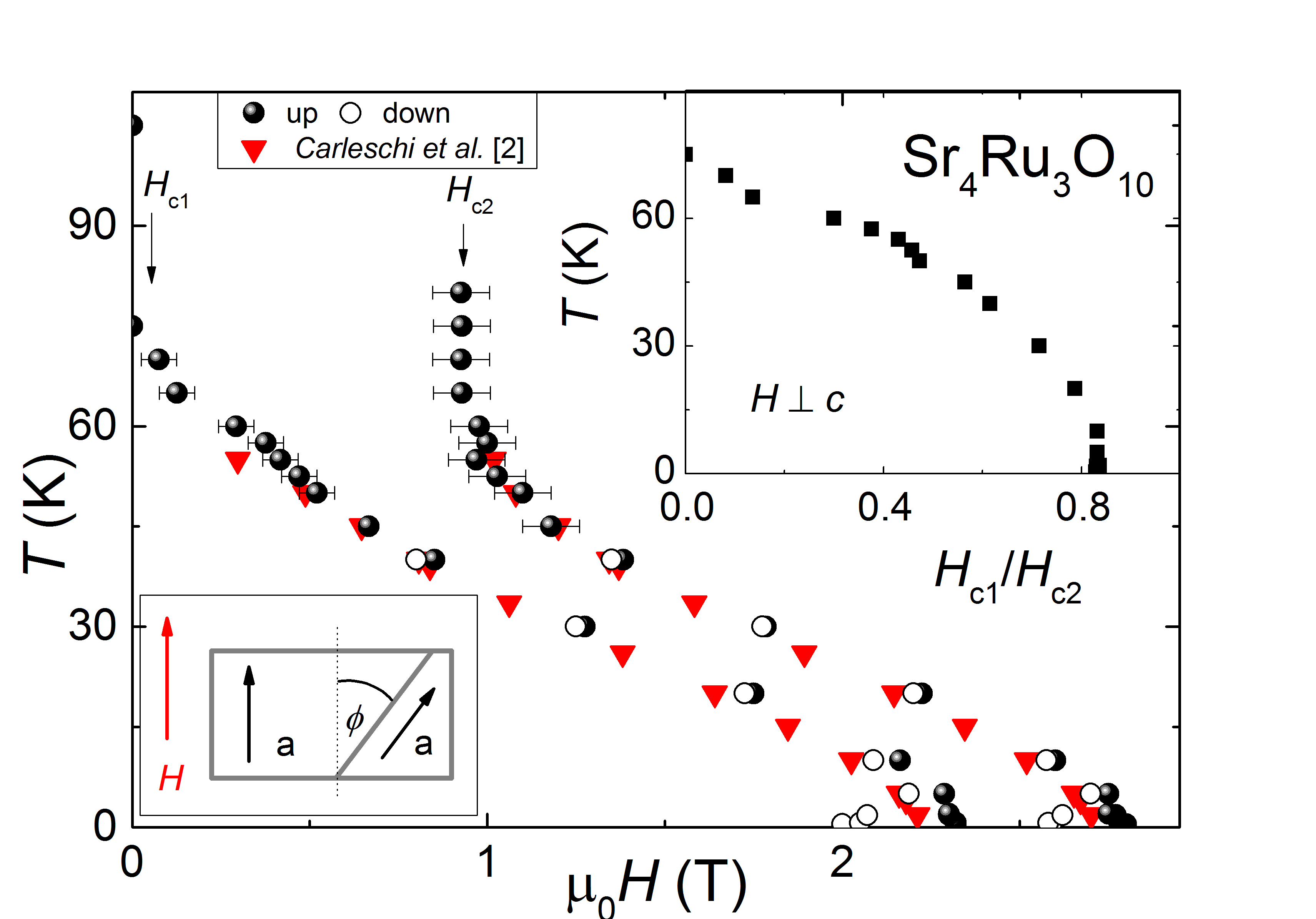}
	\caption{\label{phasedia_new}Temperature - magnetic field ($T-H$) phase diagram of \sro\ as estimated by magnetization measurements. Data of a previous study by \textit{Carleschi et al.}\cite{carleschi_14} are included (red symbols). The upper inset shows the ratio $\frac{H_{c1}}{H_{c2}}$ as a function of temperature and the lower inset a sketch of a single crystal with partial in-plane rotation (twinning).}
\end{figure}

\paragraph{Discussion}
Fig.\,\ref{phasedia_new} summarizes the magnetization results in the $H-T$ phase diagram. We add data points obtained by \textit{Carleschi et al.} \cite{carleschi_14} below 60\,K. The error bars of the critical fields reflect the discussed variation on the respective background subtraction in the intermediate temperature range. 
Striking in the phase diagram is that $H_{c1}$ is a closed phase boundary connecting the temperature with the field axis. In contrast, $H_{c2}$ is an open line starting at 2.8\,T for lowest temperatures and ending in a critical end point around 80\,K. 

One explanation for the double step in $M(H)$ at the MM transition could be that the measurements were done on a single crystal consisting of two different domains that are rotated in the $ab$-plane with respect to each other as sketched in Fig.\,\ref{phasedia_new}, lower inset. This twinning under an angle $\phi$ could explain two different values for the critical field, because the rotated sample fraction sees a smaller effective applied field $H_{eff} = H \cos \phi$. Foregoing assumption implies that the ratio $\frac{Hc1}{Hc2}(T) = \cos \phi$ does not change with temperature. The upper inset of Fig.\,\ref{phasedia_new} displays the temperature evolution of $\frac{Hc1}{Hc2}(T)$. It clearly changes from 0.8 at low temperature to zero at 80\,K where both MM transitions vanish. We interpret this temperature dependence as strong indication that the double step in the magnetization at the MM transition is an intrinsic property of \sro\ and not due to single crystal twinning or other geometrical effects.

\paragraph{Acknowledgment}
F.W. and M.J. acknowledge support from the US NSF, Division of Material Research, Grant 1157490. L.C. and B.M. were supported by the US DOE, Office of Science, Basic Energy Sciences, Materials Sciences and Engineering Division. A.M.S. acknowledges financial support from the SA-NRF (93549) and from the UJ Research Committee.

%\section{Bibliography styles}

%There are various bibliography styles available. You can select the style of your choice in the preamble of this document. These styles are Elsevier %styles based on standard styles like Harvard and Vancouver. Please use Bib\TeX\ to generate your bibliography and include DOIs whenever available.

\section*{References}

\bibliography{bib/article_201706}

\begin{thebibliography}{1}
\expandafter\ifx\csname url\endcsname\relax
  \def\url#1{\texttt{#1}}\fi
\expandafter\ifx\csname urlprefix\endcsname\relax\def\urlprefix{URL }\fi
\expandafter\ifx\csname href\endcsname\relax
  \def\href#1#2{#2} \def\path#1{#1}\fi

\bibitem{crawford_02}
M.~K. Crawford, R.~L. Harlow, W.~Marshall, Z.~Li, G.~Cao, R.~L. Lindstrom,
  Q.~Huang, J.~W. Lynn, Structure and magnetism of single crystal
  sr$_{4}$ru$_{3}$o$_{10}$: A ferromagnetic triple-layer ruthenate, Physical
  Review B 65~(21) (2002) 214412--1--5.

\bibitem{carleschi_14}
E.~Carleschi, B.~P. Doyle, R.~Fittipaldi, V.~Granata, A.~M. Strydom, M.~Cuoco,
  A.~Vecchione, Double metamagnetic transition in sr$_{4}$ru$_{3}$o$_{10}$,
  Physical Review B 90 (2014) 205120--1--9.

\bibitem{weickert_17}
F.~Weickert, L.~Civale, B.~Maiorov, M.~Jaime, M.~B. Salamon, E.~Carleschi,
  A.~M. Strydom, R.~Fittipaldi, V.~Granata, A.~Vecchione, Missing magnetism in
  sr$_{4}$ru$_{3}$o$_{10}$: Indication for antisymmetric exchange interaction,
  Scientific Reports 7 (2017) 3867.
\newblock \href {http://dx.doi.org/10.1038/s41598-017-03648-2}
  {\path{doi:10.1038/s41598-017-03648-2}}.

\bibitem{fobes_07}
D.~Fobes, M.~H. Yu, M.~Zhou, J.~Hooper, C.~J. O'Connor, M.~Rosario, Z.~Q. Mao,
  Phase diagram of the electronic states of trilayered ruthenate
  sr$_{4}$ru$_{3}$o$_{10}$, Physical Review B 75 (2007) 094429--1--5.

\bibitem{xu_07}
Z.~Xu, X.~Xu, R.~S. Freitas, Z.~Long, M.~Zhou, D.~Fobes, M.~Fang, P.~Schiffer,
  Z.~Mao, Y.~Liu, Magnetic, electrical transport, and thermoelectric properties
  of sr$_{4}$ru$_{3}$o$_{10}$: Evidence for a field-induced electronic phase
  transition at low temperatures, Physical Review B 76~(9) (2007) 094405--1--6.

\bibitem{fobes_10}
D.~Fobes, T.~J. Liu, Z.~Qu, M.~Zhou, J.~Hooper, M.~Salamon, Z.~Q. Mao,
  Anisotropy of magnetoresistance in sr$_{4}$ru$_{3}$o$_{10}$: Evidence for an
  orbital-selective metamagnetic transition, Physical Review B 81 (2010)
  172402--1--4.

\bibitem{cao_07}
G.~Cao, S.~Chikara, J.~W. Brill, P.~Schlottmann, Anomalous itinerant magnetism
  in single crystal sr$_{4}$ru$_{3}$o$_{10}$: A thermodynamic and transport
  investigation, Physical Review B 75 (2007) 024429--1--6.

\bibitem{gupta_06}
R.~Gupta, M.~Kim, H.~Barath, S.~L. Cooper, G.~Cao, Field- and pressure-induced
  phases in sr$_{4}$ru$_{3}$o$_{10}$: A spectroscopic investigation, Physical
  Review Letters 96 (2006) 067004--1--4.

\bibitem{schottenhamel_16}
W.~Schottenhamel, M.~Abdel-Hafiez, R.~Fittipaldi, V.~Granata, A.~Vecchione,
  M.~H\"ucker, A.~U.~B. Wolter, B.~B\"uchner,
  \href{http://link.aps.org/doi/10.1103/PhysRevB.94.155154}{Dilatometric study
  of the metamagnetic and ferromagnetic phases in the triple-layered
  ${\mathrm{sr}}_{4}{\mathrm{ru}}_{3}{\mathrm{o}}_{10}$ system}, Phys. Rev. B
  94 (2016) 155154.
\newblock \href {http://dx.doi.org/10.1103/PhysRevB.94.155154}
  {\path{doi:10.1103/PhysRevB.94.155154}}.
\newline\urlprefix\url{http://link.aps.org/doi/10.1103/PhysRevB.94.155154}

\end{thebibliography}

\end{document}